\documentclass[pra, twocolumn,superscriptaddress]{revtex4}

\usepackage{graphicx,float}
\usepackage{epsfig}
\usepackage{dcolumn}
\usepackage{ulem}
\usepackage{bm}
\usepackage{amsmath}
\usepackage{amssymb}
\usepackage{hyperref}
\hypersetup{colorlinks=true,linkcolor=blue,citecolor=blue,urlcolor=black}

\newcommand{\ket}[1]{\ensuremath{|\,{#1}\,\rangle}}

\begin{document}

\title{Long-distance distribution of genuine energy-time entanglement}
\date{\today}

\author{A.~Cuevas}\thanks{These authors contributed equally to this work}
\author{G.~Carvacho}\thanks{These authors contributed equally to this work}
\affiliation{Departamento de F\'isica, Universidad de Concepci\'on, 160-C Concepci\'on, Chile}
\affiliation{Center for Optics and Photonics, Universidad de Concepci\'on, Concepci\'on, Chile}
\affiliation{MSI-Nucleus for Advanced Optics, Universidad de Concepci\'on, Concepci\'on, Chile}

\author{G.~Saavedra}\thanks{These authors contributed equally to this work}
\affiliation{Center for Optics and Photonics, Universidad de Concepci\'on, Concepci\'on, Chile}
\affiliation{MSI-Nucleus for Advanced Optics, Universidad de Concepci\'on, Concepci\'on, Chile}
\affiliation{Departamento de Ingenier\'ia El\'ectrica, Universidad de Concepci\'on, 160-C Concepci\'on, Chile}

\author{J.~Cari\~ne}
\affiliation{Center for Optics and Photonics, Universidad de Concepci\'on, Concepci\'on, Chile}
\affiliation{Departamento de Ingenier\'ia El\'ectrica, Universidad de Concepci\'on, 160-C Concepci\'on, Chile}

\author{W.~A.~T.~Nogueira}
\affiliation{Departamento de F\'isica, Universidad de Concepci\'on, 160-C Concepci\'on, Chile}
\affiliation{Center for Optics and Photonics, Universidad de Concepci\'on, Concepci\'on, Chile}
\affiliation{MSI-Nucleus for Advanced Optics, Universidad de Concepci\'on, Concepci\'on, Chile}

\author{M.~Figueroa}
\affiliation{Center for Optics and Photonics, Universidad de Concepci\'on, Concepci\'on, Chile}
\affiliation{Departamento de Ingenier\'ia El\'ectrica, Universidad de Concepci\'on, 160-C Concepci\'on, Chile}

\author{A.~Cabello}
\affiliation{Departamento de F\`isica Aplicada II, Universidad de Sevilla, E-41012, Sevilla, Spain.}

\author{P.~Mataloni}
\affiliation{Dipartimento de Fisica, Sapienza Universit\`a di Roma, Piazzale Aldo Moro 5, Roma I-00185, Italy.}
\affiliation{Istituto Nazionale di Ottica (INO-CNR), Largo E. Fermi 6, I-50125 Firenze, Italy}

\author{G.~Lima}
\affiliation{Departamento de F\'isica, Universidad de Concepci\'on, 160-C Concepci\'on, Chile}
\affiliation{Center for Optics and Photonics, Universidad de Concepci\'on, Concepci\'on, Chile}
\affiliation{MSI-Nucleus for Advanced Optics, Universidad de Concepci\'on, Concepci\'on, Chile}

\author{G.~B.~Xavier}
\email{gxavier@udec.cl}
\affiliation{Center for Optics and Photonics, Universidad de Concepci\'on, Concepci\'on, Chile}
\affiliation{MSI-Nucleus for Advanced Optics, Universidad de Concepci\'on, Concepci\'on, Chile}
\affiliation{Departamento de Ingenier\'ia El\'ectrica, Universidad de Concepci\'on, 160-C Concepci\'on, Chile}

\begin{abstract}
Any practical realization of entanglement-based quantum communication must be intrinsically secure and able to span long distances avoiding the need of a straight line between the communicating parties. The violation of Bell's inequality offers a method for the certification of quantum links without knowing the inner workings of the devices. Energy-time entanglement quantum communication satisfies all these requirements. However, currently there is a fundamental obstacle with the standard configuration adopted: an intrinsic geometrical loophole that can be exploited to break the security of the communication, in addition to other loopholes. Here we show the first experimental Bell violation with energy-time entanglement distributed over 1 km of optical fibers that is free of this geometrical loophole. This is achieved by adopting a new experimental design, and by using an actively stabilized fiber-based long interferometer. Our results represent an important step towards long-distance secure quantum communication in optical fibers.
\end{abstract}

\maketitle


A fundamental application of quantum mechanics is in the development of communication systems with intrinsic unbreakable security \cite{GisinRMP}. Although quantum key distribution (QKD) is theoretically unconditionally secure, it has been experimentally demonstrated that QKD with realistic devices is prone to hacking \cite{Lydersen_2010}. A definitive solution to these practical attacks depends on the experimental violation of a Bell inequality, allowing communication with security guaranteed by the impossibility of signaling at superluminal speeds \cite{Bell,Ekert91,Kent05,Acin07}. The security can only be guaranteed if a loophole-free Bell test is performed, a major experimental task that still has not been demonstrated. With single-photons, the locality and the detection loopholes have been individually closed in separate experiments \cite{Zeilinger98, Zeilinger13, Kwiat13}. Recent experimental demonstrations of Einstein-Podolsky-Rosen (EPR) steering free of the detection loophole, also constitute an important step towards secure communications \cite{Smith2012, Bennet2012, Wittmann2012}. Progress has also been made regarding more subtle issues, such as the coincidence-time loophole \cite{Larsson2004,Kwiat13,ZeilingerReply,Larsson2013}.

Secure communication requires distributing quantum entanglement over long distances. The most common method to do it in optical fibers is based on Franson's configuration \cite{Franson89}. In the Franson scheme, each of two simultaneously emitted photons is injected into an unbalanced interferometer designed such that the uncertainty in the time of emission makes indistinguishable the two alternative paths each photon can take. Many experiments have been performed using this configuration \cite{Rarity_4km_1994, Gisin_10km_1998, Brendel99, Gisin_QKD_2000, Gisin_50km_2004, Salart2008, Salart2008b}, due to energy-time's innate robustness to decoherence in optical fibers. The problem of this method is that Franson's configuration has an intrinsic geometrical loophole and, therefore, cannot rule out all possible local explanations for the apparent violation of the Bell Clauser-Horne-Shimony-Holt (CHSH) inequality \cite{CHSH69, Aerts_PRL_1999, Cabello_genuine_2009}. Current experiments based on Franson's scheme only rule out local models making additional assumptions \cite{Franson99, Franson09, Larsson10}. On the practical side, this loophole can be exploited by eavesdroppers to break the security of the communication (see, e.g., the Trojan horse attack to Franson-based quantum cryptography in Ref. \cite{Larsson02}). Some solutions to this problem have been proposed. One consists of keeping Franson's configuration but replacing passive by active switchers \cite{Brendel99}. However, this solution has never been implemented in an experiment. Another solution consists of replacing Franson's interferometers by unbalanced cross-linked interferometers wrapped around the photon-pair source in a ``hug'' configuration introduced in \cite{Cabello_genuine_2009}. It has been recently demonstrated in table-top experiments \cite{Glima_genuine_2010,Vallone_genuine_2011}. Going to larger distances was thought to be unfeasible unless costly stabilization systems used for large gravitational wave detectors were applied \cite{Cabello_genuine_2009}.

Here we report the first experimental violation of the Bell CHSH inequality with genuine energy-time entanglement (i.e., free of the geometrical loophole) distributed through more than 1 km of optical fibers. The energy-time entangled photon-pair source is placed close to Alice, with one of the photons propagating through a short bulk optics unbalanced Mach-Zenhder (UMZ) interferometer. The other photon is transmitted through the second UMZ interferometer of the hug configuration, which is composed of 1-km long fiber optical arms. We show that a Bell violation can be obtained with a home-made active phase stabilization system, demonstrating that the hug configuration is indeed practical for long-distance entanglement distribution. The results presented here have major implications for secure quantum communication between parties that are not in the same straight line, opening up a path towards a loophole-free Bell test with energy-time entanglement. We stress that even if the detection and locality loopholes are simultaneously closed in a single experiment, energy-time entanglement will still remain unusable as a resource for device-independent quantum communication if the geometrical loophole is not addressed.

\begin{figure*}[thb]
\centerline{\includegraphics[width=0.9\textwidth]{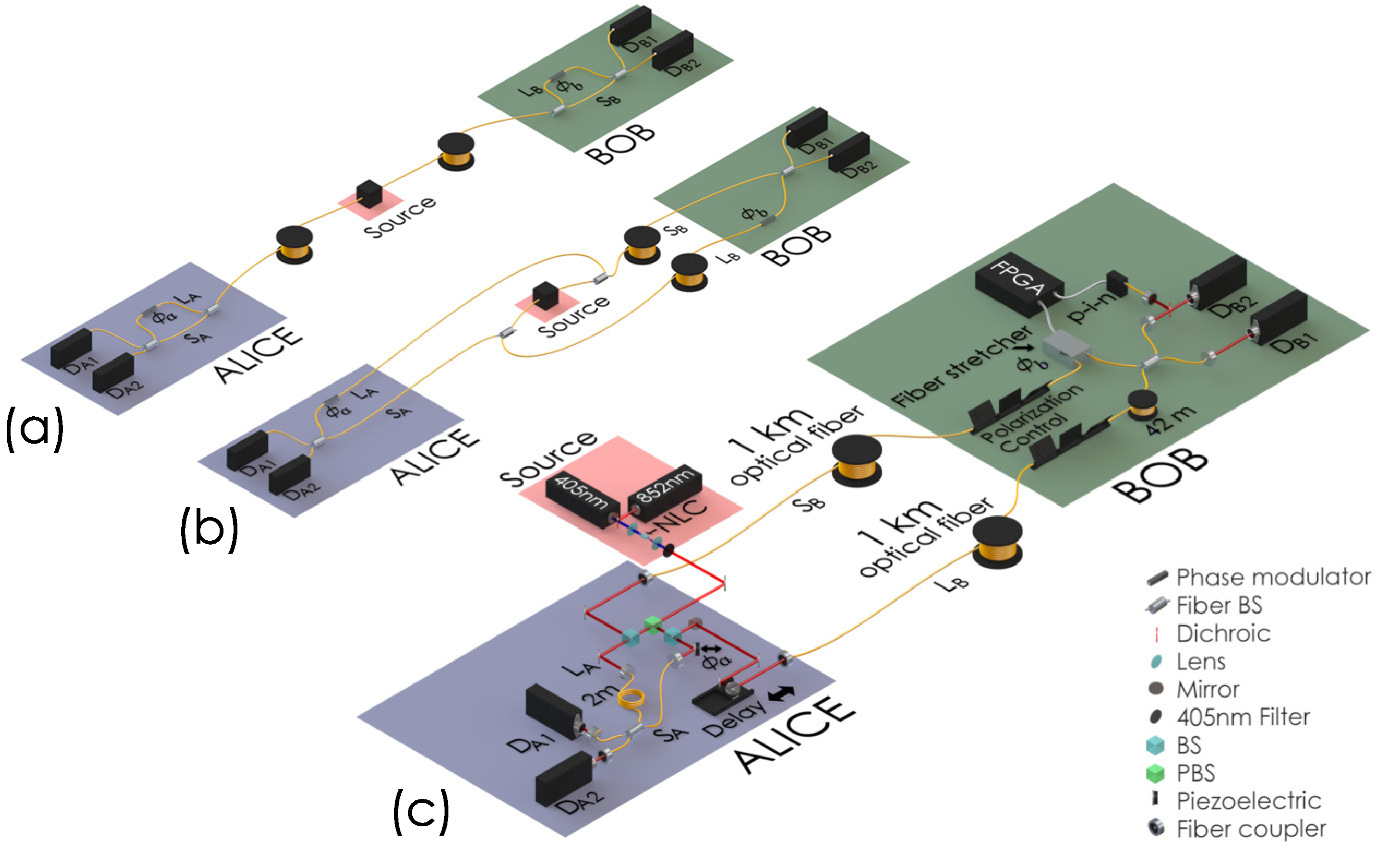}}
\caption{Energy-time Bell test configurations. (a) Typical fiber-based implementation of Franson's scheme \cite{Franson89,Gisin_QKD_2000}. (b) Fiber-based version of the hug configuration \cite{Cabello_genuine_2009}. (c) Experimental setup used for implementing the hug configuration with a long fiber-based interferometer. Photon pairs generated in a non-linear crystal (NLC) are sent through two cross-linked unbalanced Mach-Zehnder interferometers at Alice and Bob's sites. An optical delay line is used to set the indistinguishability between the short-short and long-long two-photon paths (see the main text for details). The long interferometer is comprised of a 1-km long telecom single-mode fiber spool in each arm. An extra 40 m of optical fiber is added in the $S_{\text{B}}$ arm by the fiber stretcher, which is then balanced by an equal amount of fiber in the other arm. The total arm length in this interferometer is then 1.04 km. The long-short path length difference in both interferometers is 2 m of optical fibers.} \label{fig1}
\end{figure*}

\section*{Results}
\subsubsection*{Experimental setup}

The extensively employed Franson configuration is shown schematically in Fig. 1(a). It consists of two UMZ interferometers with the two arms in each one defined as the short (S) and long (L) paths respectively. In order to avoid single-photon interference, the long-short path difference is made to be much larger than the single-photon coherence length. The issue lies with the fact that the coincident detection events are post-selected for the verification of the Bell violation, and since this post-selection is local, it gives rise to hidden variable models \cite{Aerts_PRL_1999, Cabello_genuine_2009}. In the hug configuration, shown in Fig. 1(b), the unbalanced interferometers are formed shortly after each output mode of the source using beamsplitters, with the difference that the arms are crossed around the source. This geometrical property assures that the short-long or long-short events can only be routed to either Alice or Bob detectors, effectively removing the need of local post-selections.

The experimental setup used is depicted in Fig. 1(c). Degenerate 806 nm photon pairs with orthogonal polarizations are produced from spontaneous parametric down-conversion in a bulk non-linear periodically poled potassium titanyl phosphate (KTP) 20 mm long crystal. The crystal is pumped by a single-longitudinal mode continuous wave (CW) 403 nm laser with a coherence length greater than 20 m, and with 2.2 mW of optical power. A long coherence pump laser is necessary to create a fundamental uncertainty in the emission time of the down-converted photons, and thus set up as indistinguishable, the possible paths they can take in both (equally) unbalanced interferometers \cite{Cabello_genuine_2009}. When this condition is satisfied, the Bell state $\ket{\Phi^+}=\frac{1}{\sqrt{2}}(\ket{SS}+\ket{LL})$ is generated, where $S$ and $L$ indicate the short and long arms, respectively. The pump laser is weakly focused in the crystal in order to obtain a high collection efficiency in the optical fibers \cite{Ljunggren_PRA_fibers}. The emitted photons are deterministically split by a polarizing beam splitter (PBS), with each output mode sent to a beamsplitter (BS). The two cross-linked interferometers required in the hug configuration are built on these four modes connecting the source to Alice and Bob, $S_{\text{i}}$ and $L_{\text{i}}$, with $i=\text{A,B}$. The two modes $S_\text{{A}}$ and $L_\text{{A}}$ form a short-distance UMZ interferometer. Mode $S_\text{{A}}$ is reflected on a piezo-electric mounted mirror responsible for generating the phase shift $\phi_{\text{a}}$ employed by Alice. The final beamsplitter is a single-mode fiber-based component to optimize the overlap of the transverse spatial modes. The long-short path length difference is 2 m. This was chosen such that the time difference is larger than the coincidence temporal window (4 ns). The length of the short arm is approximately 1 m.

The other two modes, $S_{\text{B}}$ and $L_{\text{B}}$ connect the source to Bob through a long UMZ fiber-based interferometer. Its arms are built using 1-km long telecom spooled single-mode fibers. One spool at each arm. The use of telecom fibers is to demonstrate that the distribution of genuine energy-time entanglement is compatible with the already installed optical-based world-wide network. Since multi-mode propagation occurs in the telecom fibers at 806 nm, the arms are connected to a 780 nm single-mode fiber-optical beamsplitter at Bob's station, performing transverse spatial mode filtering \cite{Jennewein_APL_2010}. This interferometer presents the same 2 m difference between the long and short arms, as in Alice's case. For the active phase stabilization a piezo-electric fiber stretcher (FS) with 40 m of wound fiber is used in the $S_B$ arm of Bob's interferometer. Therefore an equal amount of fiber length must be added to the $L_B$ arm for balancing purposes. Thus, the total arm length in this interferometer is 1.04 km. With the exception of the 1 km fiber spools, all other fiber-optic components are single-mode at 780 nm and above. The polarization is adjusted in both arms of the long interferometer with polarization controllers to avoid which-way path information.

\begin{figure}[t]
\centerline{\includegraphics[width=0.5\textwidth]{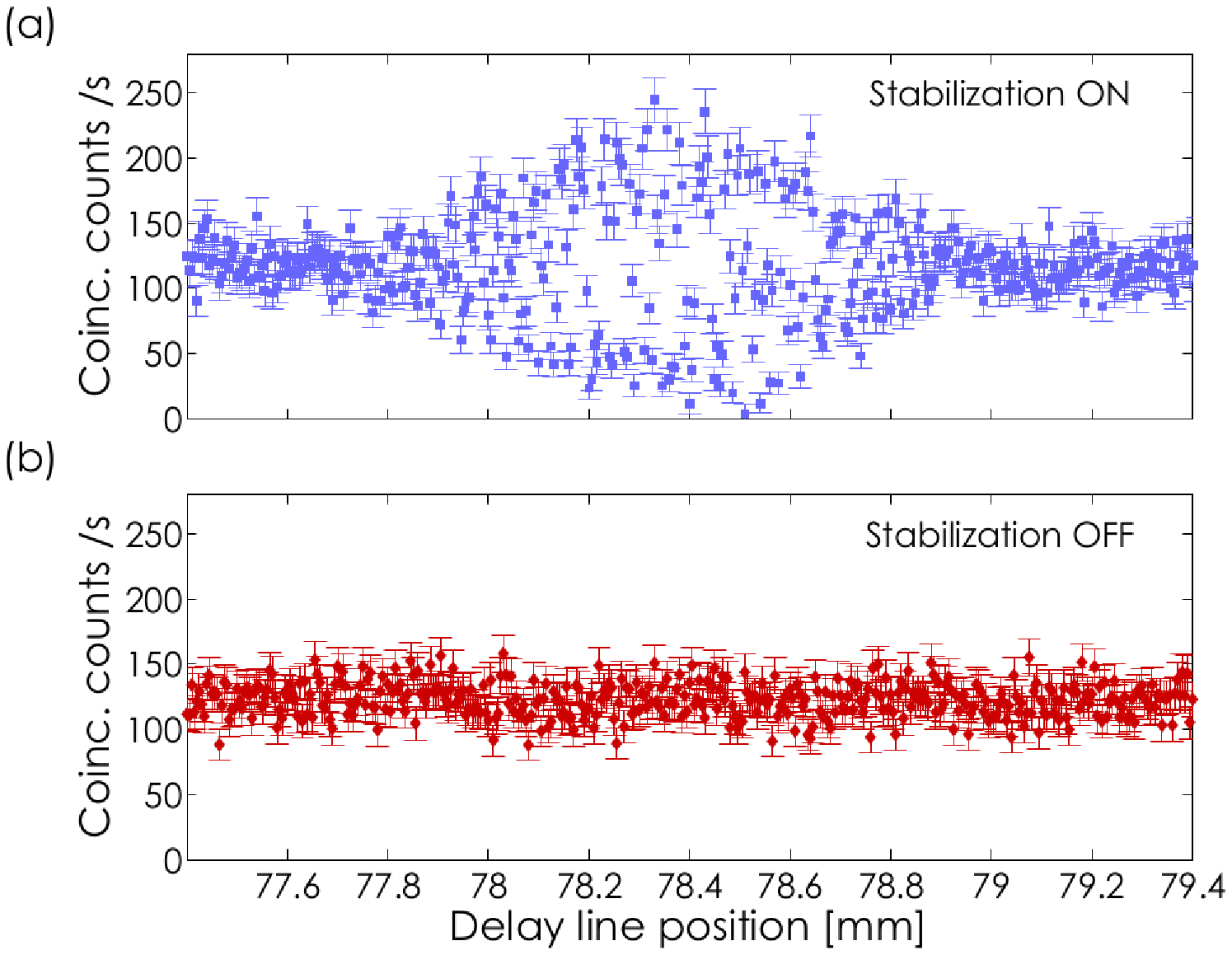}}
\caption{Net coincident counts versus the delay line position. A two-photon interference pattern is observed in case (a) with the stabilization system active, while (b) shows a second measurement over the same range of the delay line with stabilization turned off. The error bars shows the standard deviation assuming a poissonian distribution for the photon statistics.} \label{fig2}
\end{figure}

\subsubsection*{Active phase stabilization}

A major experimental challenge in this implementation is the compensation of random phase fluctuations caused by the environment in long interferometers. This problem is successfully solved here for the first time by adapting to the hug configuration some ideas developed for different purposes \cite{guix_OL}, providing an affordable alternative to the costly stabilization systems suggested in Ref. \cite{Cabello_genuine_2009}. The solution is achieved by injecting a second laser into the system to provide real-time feedback information for the field programmable gate array (FPGA) based control electronics. As mentioned above, a piezo-electric FS composed of 40 m of wound fiber is used for the active phase stabilization of the long fiber-based interferometer. This device allows for a fiber expansion of up to 5 mm at the maximum driving voltage. This is essential as it allows several wavelengths of phase drift to be compensated in a long interferometer. The FS is driven by the control electronics in response to the environmentally-induced phase drifts imposed on the feedback optical signal when propagating in the interferometer. The control system enables phase-locking of the interferometer by using a custom designed closed-loop proportional-integral-derivative (PID) control algorithm. Additional details are provided in Methods section. A bulk-optics delay line, with a movement range of 150 mm, is used to set $L_{\text{A}}-S_{\text{A}} = L_{\text{B}}-S_{\text{B}}$ (see also Methods for additional details). To demonstrate the importance of the active stabilization in the experiment, we show in Fig. 2(a) the net coincidence counts/s between detectors $D_{\text{A1}}$ and $D_{\text{B1}}$ as a function of the delay line position around the indistinguishability point. The stabilization system was kept active maintaining $\phi_{\text{b}}$ constant while the delay line was moving. In Alice's side, the piezo-mounted mirror was driven during the measurement with a slowly varying voltage ramp, therefore modulating $\phi_{\text{a}}$. This is to ensure that two-photon interference fringes are observed in the indistinguishability region. The two-photon interference pattern in Fig. 2(a) is clearly observed. The crucial point is that two-photon interference can not be observed without active phase stabilization due to the rapidly and randomly varying phase drift in a long interferometer. This is shown in Fig. 2(b), where the scan in the same delay is performed with the control system turned off and all other settings kept the same as in Fig. 2(a).

\begin{figure}[h]
\centerline{\includegraphics[width=0.5\textwidth]{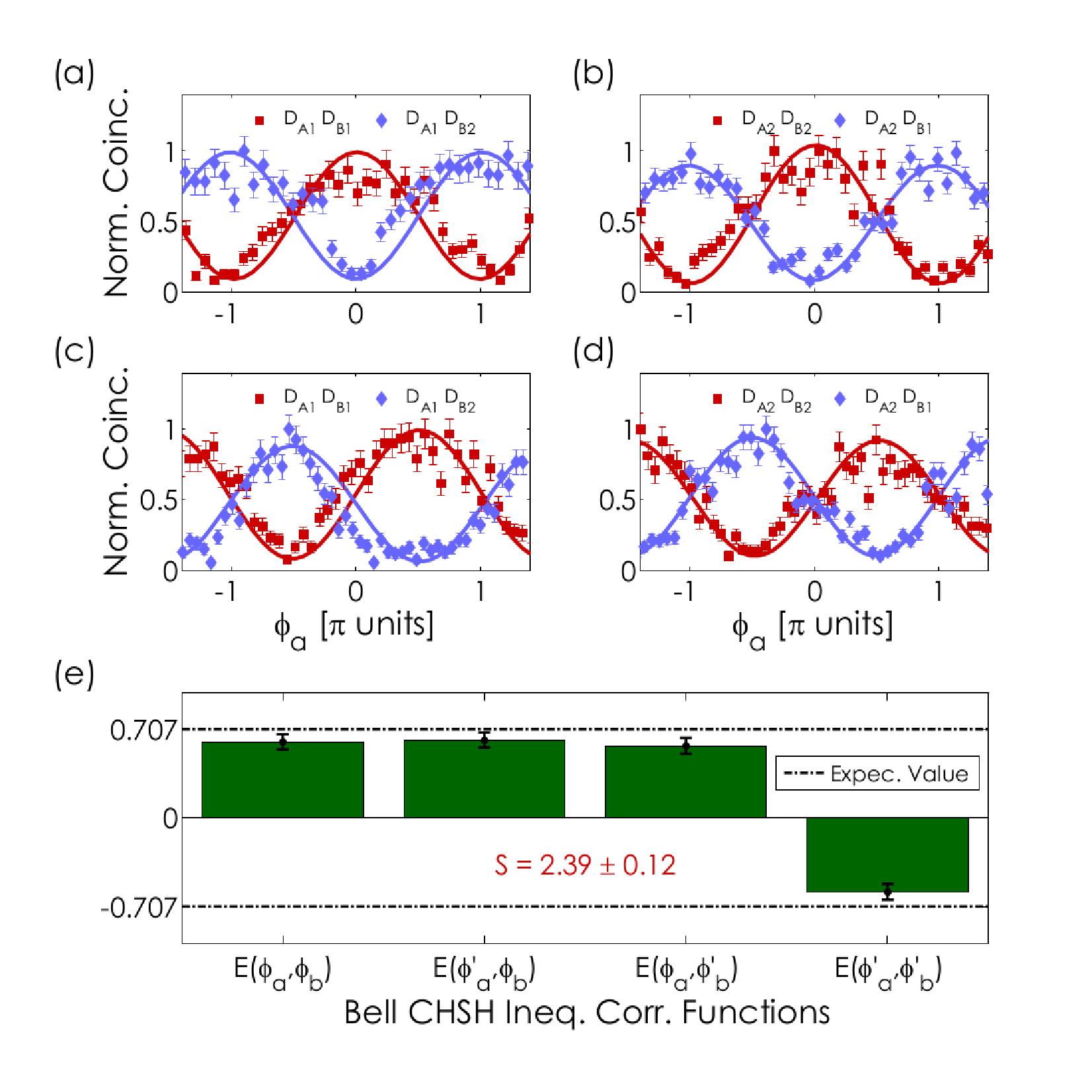}}
\vspace{-0.5cm}
\caption{Coincidence interference curves and Bell CHSH inequality violation. Graphs (a)-(d) show the normalized coincidence detections across Alice and Bob's detectors without accidental subtraction. (a) and (b) correspond to the cases where $\phi_{\text{b}}=0$ and (c) and (d) to the cases where $\phi_{\text{b}}=\pi/2$. Integration time for each data point is 1 s with an average rate of $\approx 100$ coincidences counts. The rate of single counts is $\approx$ 74000 and 54000 detections per second for each detector of Alice and Bob respectively. The curves are obtained varying the phase difference $\phi_{\text{a}}$ in Alice's interferometer with the moving mirror, while actively keeping fixed the relative phase $\phi_{\text{b}}$ . (e) shows the measured value for each probability correlation function ($E$) appearing in the Bell CHSH inequality. Solid black lines show the maximum possible value for each function $E$. The corresponding experimental Bell violation yields $S = 2.39 \pm 0.12$. The difference between the fit curve and the data points in (a)--(d) is due to random drifts in the piezo movement and the statistical distribution of the single-photon detections. The error bars shows the standard deviation assuming a poissonian distribution for the photon statistics.} \label{fig3}
\end{figure}

\subsubsection*{Long-distance Bell inequality violation}

The violation of the Bell CHSH inequality was performed with the delay line set at the center point of the two-photon interference pattern, the active phase control is kept active and the piezo-mounted mirror once again slowly modulated (this time with a $\sim$ 30 s period). The Bell CHSH inequality is defined through the expression $S = E(\phi_{\text{a}},\phi_{\text{b}}) + E(\phi^{\prime}_{\text{a}},\phi_{\text{b}}) + E(\phi_{\text{a}},\phi^{\prime}_{\text{b}}) - E(\phi^{\prime}_{\text{a}},\phi^{\prime}_{\text{b}})$, where $E(\phi_{\text{a}}, \phi_{\text{b}}) = P_{11}(\phi_{\text{a}}, \phi_{\text{b}}) + P_{22}(\phi_{\text{a}}, \phi_{\text{b}}) - P_{12}(\phi_{\text{a}}, \phi_{\text{b}}) - P_{21}(\phi_{\text{a}}, \phi_{\text{b}})$, with $P_{\text{ij}}(\phi_{\text{a}}, \phi_{\text{b}})$ corresponding to the probability of a coincident detection at Alice and Bob's detectors $i$ and $j$ respectively, while the relative phases $\phi_{\text{a}}$ and $\phi_{\text{b}}$ are applied to Alice and Bob's interferometers. For the maximum violation of the Bell CHSH inequality the phase settings are $\phi_{\text{a}} = \pi/4$, $\phi_{\text{b}} = 0$, $\phi_{\text{a}}^{\prime} = -\pi/4$ and $\phi_{\text{b}}^{\prime} = \pi/2$ \cite{CHSH69, Glima_genuine_2010}. The phase control system is used to switch between Bob's two phase settings, 0 and $\pi/2$. The eight measured curves across all output combinations are displayed in Fig. 3(a)-(d). The measured average raw visibilitiy is (84.36 $\pm$ 0.47)\%. When the accidental coincidences are subtracted, the net visibility rises to (95.12 $\pm$ 0.20)\%. The recorded values of the probability correlation functions ($E$) for the case with no accidental subtraction are shown in Fig. 3(e). The corresponding violation of the Bell CHSH inequality in this case yields $S = 2.39 \pm 0.12$, surpassing the classical limit by 3.25 standard deviations. Our experimental results are comparable to previous long-distance Bell experiments using energy-time entanglement based on the Franson configuration \cite{Gisin_10km_1998}.

\section*{Discussion}

Future quantum communication systems must work over long distances, avoid the need of a straight line between the communicating parties and fit within existing communication infrastructures. In addition, security must be based on physical principles rather than on unproven assumptions. For these reasons, energy-time entanglement-based quantum communication is, in principle, the ideal solution. However, the fact that standard setups for creating energy-time entanglement are intrinsically insecure even assuming perfect detection efficiency constitutes a fundamental hurdle. Here we have demonstrated for the first time that genuine energy-time entanglement can be distributed over long distances, thus removing one fundamental obstacle for practical and secure quantum communication with optical fibers. Note that even though polarization entanglement has also been distributed through optical fibers \cite{Sauge2007, Hubel2007}, energy-time entanglement has the advantage of having an innate robustness to decoherence in optical fibers. 

In commercial networks, remote nodes are connected with optical cables composed of many optical fibers. For an experiment, based on the hug configuration and with spatial separation between Alice and Bob, two fibers of the same cable can be used. In this case, as the phase drift acts almost equally on both fibers, it is very likely that the scheme adopted here for the stabilization of the fiber-based long interferometer can also be used. Further investigations are required in this direction. Nevertheless, our work is a first step towards practical long distance secure quantum communication based on energy-time entanglement, as it fixes the previous security issues with all other demonstrations using Franson's configuration.

\subsection*{Methods}

\subsubsection*{Stabilization system}

A long-coherence single-longitudinal mode CW laser operating at 852 nm is used as a feedback optical signal. The feedback optical signal is combined with the pump beam on a dichroic mirror before the crystal. The FS is a commercial off-the-shelf device consisting of a 40 m long optical fiber spooled around a piezo-electric element. The feedback optical signal is detected by an amplified p-i-n photodetector in one of the outputs of the long interferometer, after being split by a dichroic mirror. In order to avoid unwanted noise generated from the control laser, three extra optical filters are employed: one bandpass 850 nm optical filter (10 nm of full width at half-maximum -- FWHM), placed at the output of the 852 nm laser, and two bandpass 806 nm filters in series (5 nm of FWHM) are inserted before each single-photon detector. For the sake of clarity these are not shown in Fig. 1(c).

The phase setting $\phi_{\text{b}}$ applied by Bob was controlled using the set-point in the FPGA electronics. It reads the optical intensity of the feedback signal, which gives information on the current relative phase between the interferometer arms. Furthermore it calculates the derivative of the signal, to remove ambiguity arising from the phase information. With both the intensity of the signal and the derivative, the control is able to fix any constant relative phase difference between the interferometer arms. The total bandwidth of the control system including the optical components (stretcher and detector) is of approximately 5 kHz.

\subsubsection*{Indistinguishability between the two-photon paths}

To generate the indistinguishability between the $\ket{SS}$ and $\ket{LL}$ paths, a bulk-optics delay line, with a movement range of 150 mm, is used. Since it is experimentally challenging to properly balance two long paths within the two-photon coherence length ($\approx 1$ mm in our case), a previous adjustment of the 1.04 km arms was performed in an external interferometer with a Fabry-Perot (FP) laser source for a course adjustment. Then a broadband light source (a light emitting diode) was used replacing the FP laser to set the arm lengths to be equal within less than 1 mm. The arms were then installed in the setup, and the extra 2 m of optical fiber added in the long arm.

\subsection*{Acknowledgments}

The authors thank M. Barbieri for valuable discussions. This work was supported by the grants FONDECYT 11110115 and 1120067, CONICYT PFB08-024 and Milenio P10-030-F. A. Cuevas, G. C. and J. C. acknowledge the financial support of CONICYT, while M. F. acknowledges support of  FONDECYT 1121010. A. Cabello was also supported by Project No.\ FIS2011-29400 (MINECO, Spain).

\subsection*{Authors' Contributions}

A. Cuevas, G. C., G. S. and J. C., with assistance from W. N., M. F., P. M., G. L. and G. X., performed the experiment and analyzed the data. A. Cabello, P. M., G. L. and G. X. conceived and designed the experiment and wrote the paper. All authors agree to the contents of the paper.

\subsection*{Additional information}

Correspondence and requests for materials should be addressed to G. X.
The authors declare no competing financial interests.

\end{document}